\documentclass[11pt, prd,preprintnumbers,amsmath,amssymb, superscriptaddress]{revtex4}

\pdfoutput=1
\usepackage{latexsym}
\usepackage{wasysym}
\usepackage{amssymb}
\usepackage{cancel}
\usepackage{epsfig,amsmath,graphics}
\usepackage{epstopdf}
\usepackage{verbatim}
\usepackage[hypertex]{hyperref}
\usepackage{graphicx}
\usepackage{subfigure}
\usepackage[version=3]{mhchem}

\newcommand{\keV}{\text{ keV}}
\newcommand{\keVr}{\text{ keVr}}
\newcommand{\keVee}{\text{ keVee}}
\newcommand{\MeV}{\text{ MeV}}
\newcommand{\GeV}{\text{ GeV}}

\newcommand{\kms}{\text{ }\frac{\text{km}}{\text{s}}}

\newcommand{\OrderOne}{\mathcal{O}\left(1\right)}
\newcommand{\mchi}{m_{\chi}}
\newcommand{\mAp}{m_{A'}}
\newcommand{\Leff}{\mathcal{L}_{eff}}

\begin{document}

\title{Exothermic Dark Matter}

\date{\today}
             
 \author{Peter W. Graham}
\affiliation{Department of Physics, Stanford University, Stanford, California 94305}

\author{Roni Harnik}
\affiliation{Theoretical Physics Department, Fermilab, Batavia, IL60510, USA}

\author{Surjeet Rajendran}
\affiliation{Center for Theoretical Physics, Laboratory for Nuclear Science and Department of Physics, Massachusetts Institute of Technology, Cambridge, MA 02139, USA}

\author{Prashant Saraswat}
\affiliation{Department of Physics, Stanford University, Stanford, California 94305}

\preprint{FERMILAB-PUB-10-062-T}
\preprint{MIT-CTP 4140}

\begin{abstract}
We propose a novel mechanism for dark matter to explain the observed annual modulation signal at DAMA/LIBRA which avoids existing constraints from every other dark matter direct detection experiment including CRESST, CDMS, and XENON10.    The dark matter consists of at least two light states with mass $\sim$ few GeV and splittings $\sim 5$ keV.  It is natural for the heavier states to be cosmologically long-lived and to make up an $\OrderOne$ fraction of the dark matter.  Direct detection rates are dominated by the exothermic reactions in which an excited dark matter state down-scatters off of a nucleus, becoming a lower energy state.  In contrast to (endothermic) inelastic dark matter, the most sensitive experiments for exothermic dark matter are those with light nuclei and low threshold energies.  Interestingly, this model can also naturally account for the observed low-energy events at CoGeNT.  The only significant constraint on the model arises from the DAMA/LIBRA unmodulated spectrum but
it can be tested in the near future by a low-threshold analysis of CDMS-Si and possibly other experiments including CRESST, COUPP, and XENON100.
\end{abstract}

\maketitle

\tableofcontents

\section{Introduction}
\label{Sec:Intro}

A variety of experiments are currently probing the interaction between dark matter and the standard model. These dark matter direct detection experiments are at sensitivities where they can observe weak scale interactions between nuclei and dark matter. The DAMA collaboration has observed an annual modulation of the event rate in their NaI-target detectors at the 8.9$\sigma$ level \cite{DAMAResults}. This can be interpreted as modulation of the rate of dark matter (DM) scatters in the detector due to the changing velocity of the Earth through the DM halo. However, if one assumes an elastically scattering WIMP and standard quenching of nuclear recoils then this signal is in conflict with null results from other experiments, particularly CDMS and XENON10 \cite{Kopp:2009qt}. 

The choice of target nuclei and the technique employed to detect  dark matter is unique to each experiment.  Consequently, the comparison of results from different experiments requires the specification of an underlying model that determines the interaction between the dark matter and the nucleus. A change in this underlying model could alter the expected event rate and the nuclear energy recoil spectrum in different experiments. In particular, many  experiments are optimized to look for the elastic scattering of dark matter with nucleons. This  leads to the familiar exponentially falling nuclear recoil spectrum. However, by modifying the nature of the dark matter - nucleus interaction, this spectrum could be modified, thus changing the sensitivity of different experiments  \cite{IDMOriginal,IDMinLightofDAMA, Khlopov:2008ty}.

This strategy has been used to resolve the conflict between the observations of DAMA and the null results of other experiments. One possibility is upscattering inelastic dark matter (iDM) \cite{IDMOriginal,IDMinLightofDAMA}: dark matter of $O(100 \GeV)$ mass that scatters inelastically off nuclei to a higher mass state. For a mass splitting of $\delta \sim 100 \keV$, comparable to typical WIMP kinetic energies, this produces a nuclear recoil energy spectrum that is peaked at high energies, in contrast to the falling exponential expected from elastic scattering. Such spectra can avoid constraints from experiments that focus on low-energy recoils. 

iDM is most constrained by experiments that observe spectra at high recoil energies. The XENON10 collaboration has analyzed their data up to recoil energies of 75 keV and found no dark matter signal \cite{Xe10Results}. Additionally, iDM preferentially scatters off heavy nuclei. The null results from  CRESST-II \cite{CRESSTIICommissioning,CRESSTTalk} and ZEPLIN-III \cite{Akimov:2010vk} rule out most of the parameter space under the assumption of a standard Maxwell-Boltzmann distribution of DM halo velocities. Other halo models may relax some of these constraints and open parts of parameter space \cite{MarchRussell:2008dy,Non-Maxwellian}.

Another avenue that has been explored to explain the DAMA signal is light dark matter (LDM) \cite{Petriello:2008jj,Fairbairn:2008gz,Chang:2008xa,ModerateChanneling}. This relies on the possibility of a channeling effect in DAMA. DAMA can only measure the electronic energy (denoted by units of ``keVee") deposited in the detector by a recoiling nucleus, which is typically a small fraction of the total recoil energy (in units of ``keVr"). Some recoils however may be ``channeled" through the NaI crystal and deposit almost 100\% of their energy electronically. While the DAMA collaboration has not reported experimental measurements of the channeling effect in their detector, they have theoretically estimated it to be $\sim 30\%$ for recoil energies of $\sim 3 \keV$ \cite{DAMAChanneling}. If channeling occurs then DAMA can observe nuclear recoils at lower energies than previously anticipated. This would give DAMA a low energy threshold compared to most other experiments, allowing it to probe light ($\sim 10 \GeV$) dark matter that other experiments may be blind to. Recently it has been noted that light dark matter can also explain the excess of low-energy events observed by the CoGeNT experiment \cite{CoGeNT}, for a slightly different parameter space than the fit to DAMA \cite{Fitzpatrick:2010em,Interpretations}.   However, this scenario is severely constrained by the null results from XENON10 and CDMS Silicon, which also have low thresholds. In particular, we find that incorporation of all the CDMS Silicon datasets strongly disfavors this explanation of DAMA (see Figure \ref{Fig:ElasticMSigma}).

In this paper we explore the possibility of explaining the DAMA signal through exothermic dark matter (exoDM); i.e. dark matter that can exist in two states with a small mass splitting, just as in conventional iDM, but which scatters from an excited state to a lower state to produce the signal observed by DAMA. (Although this model may also be described as ``inelastic dark matter", we will use that term to refer to the upscattering scenario). In this exothermic process, the energy of the recoiling nucleus is peaked around a scale that is proportional to the splitting between the dark matter states and is inversely proportional to the nuclear mass. Consequently, the nuclear recoils caused by exoDM are more visible in experiments with light nuclei and low thresholds.

The approach of this model to reconciling the DAMA results with other experiments is similar to earlier proposals of light dark matter: we consider a parameter space which produces a modulation signal at DAMA while scattering below the recoil energy detection threshold in other experiments. The assumption of downscattering allows us to fit DAMA with lower mass WIMPs ($2 - 5 \GeV$) which are less constrained by other existing experiments. There is some tension in this model between fitting the DAMA modulation and not exceeding the unmodulated rates observed in the same experiment; however as we will discuss these constraints rely on assumptions about the response of the DAMA detectors at very low energies which has not been well-measured. Similarly, XENON10 could also constrain this parameter space, but these constraints also depend upon uncertainties in the very-low-energy response of XENON10. The CDMS experiment does not constrain this parameter space.  The CoGENT excess may be accounted for with the same parameters that fit DAMA if some fraction of events are channeled in germanium as well. 
 
The model-building aspects of the exoDM scenario are similar to conventional iDM. Generically,  most iDM scenarios lead to a cosmologically long lived relic population of excited states \cite{MetastableWIMPs,pospelov}. As noted previously, downscattering of such excited states with $O(100 \keV)$  splittings can produce dramatic signals at high energies in direct detection experiments \cite{MetastableWIMPs,pospelov,Lang:2010cd}, though these can strongly constrain inelastic upscattering explanations of DAMA. Our parameter space in contrast has splittings of a few keV and produces peaked signals at very low energies. Because this signal is below threshold for most experiments, we evade current bounds; however because of the very high cross sections and rates we predict striking signals for experiments with sufficiently low thresholds. 

\section{Kinematics of Downscattering}
\label{Sec:Kinematics}

\begin{figure}
\begin{center}
\includegraphics[width = 4.1 in]{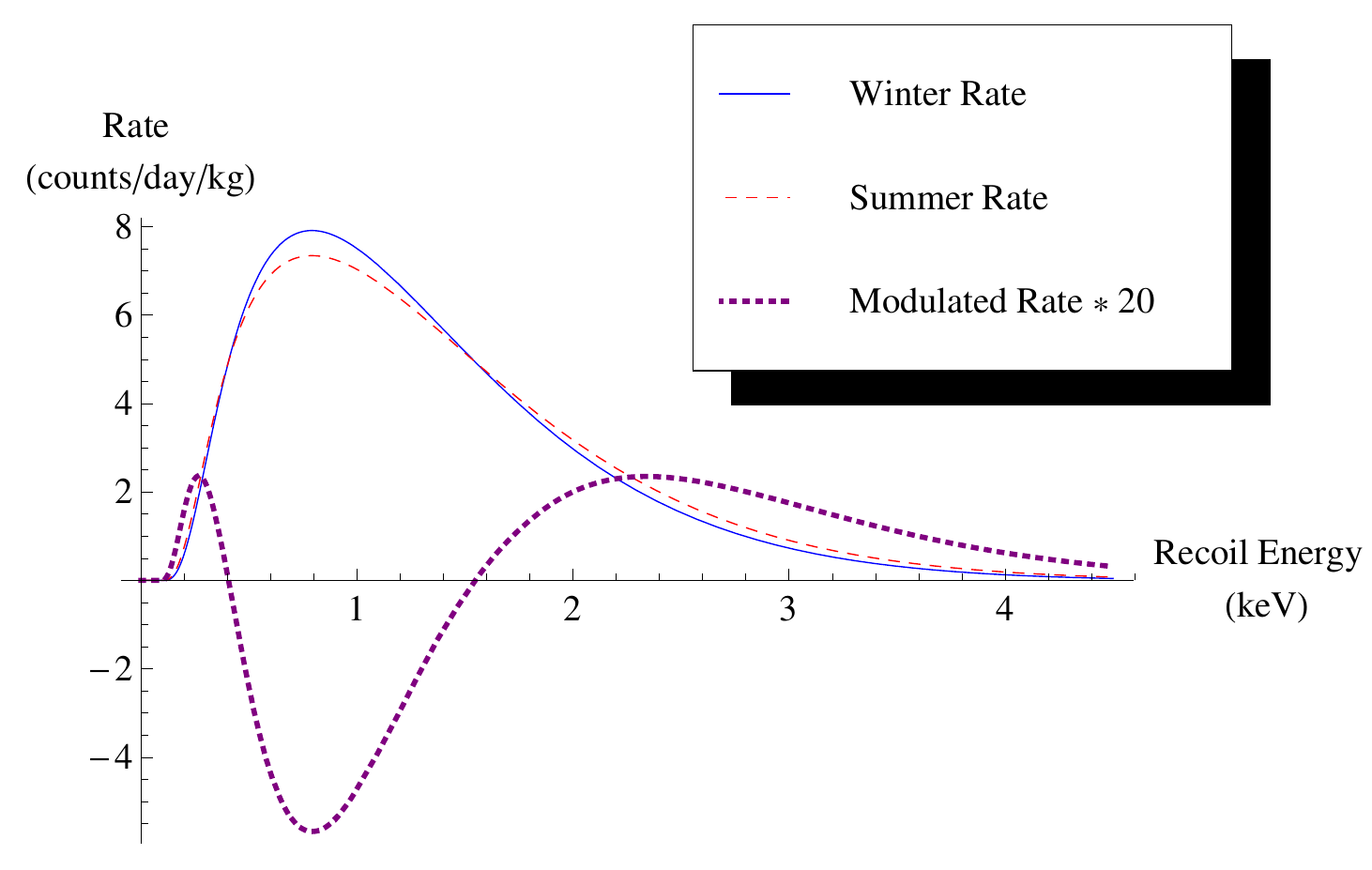}
\caption{ \label{Fig:Modulation}  Sample recoil energy spectra for downscattering off of sodium during winter when the Earth is moving with the DM halo (blue line) and during summer when it is moving against the halo (red dashed). The modulation spectrum is half of the difference between these two curves, shown here enlarged by a factor of 20 (purple dotted). Here we have taken a dark matter mass of $\mchi = 3.5 \GeV$ and a splitting of $\delta = 6 \keV$ and chosen the cross-section to fit the DAMA modulation signal. Note that the modulation is $\sim 1\%$ of the the unmodulated rate. This is tolerable for this model because the scattering is below threshold at most experiments.}
\end{center}
\end{figure}

To fit the DAMA signal (assuming some amount of channeling) we will prefer light ($2 - 4 \GeV$) dark matter downscattering with a mass splitting of $\sim 5 - 10\keV$. At first glance attempting to explain the DAMA signal with downscattering seems counterproductive: exothermic low-energy scattering rates are independent of velocity for kinetic energies less than the mass splitting, so we expect little or no modulation of the total downscattering rate.\footnotetext[1]{One may attempt to choose scattering operators so as to introduce velocity dependence into the low-energy matrix element, as considered in \cite{Bernabei:2008mv}; however this is difficult to achieve for downscattering as the momentum transfer is dominated by the mass splitting and not the initial DM kinetic energy.} However, the \emph{shape} of the exothermic recoil energy spectrum does depend on velocity, so that the rate at any given energy modulates even though the modulation is zero when integrated over energy. In particular, for light DM particles with kinetic energies much less than the splitting, exothermic spectra exhibit a peak at $E_R = \delta\frac{\mchi}{m_N}$  where $\mchi$ is the dark matter mass and $m_N$ is the nucleus mass. This peak corresponds to zero-velocity scattering where the final momentum of the dark matter particle $p_{DM} = 2 \mchi \delta $ is equal to the recoil momentum of the nucleus $p_N=2 m_N E_R$. The spread of the spectrum about this peak is proportional to the dark matter kinetic energy and thus modulates annually. Therefore the modulation spectrum (summer rate minus winter rate) is negative around the peak of the unmodulated spectrum and positive around the tails, as shown in Figure \ref{Fig:Modulation}. Because the DAMA modulation data extends from 2 keVee and above, we can fit the signal to the positive modulation region at higher energies. The ratio of this modulated rate to the total unmodulated rate is very small in this scenario ($\sim 1 \%$), so the DAMA unmodulated spectrum provides an important constraint on the model, as discussed below.

\begin{figure}
\begin{center}
\includegraphics[width = 4.1 in]{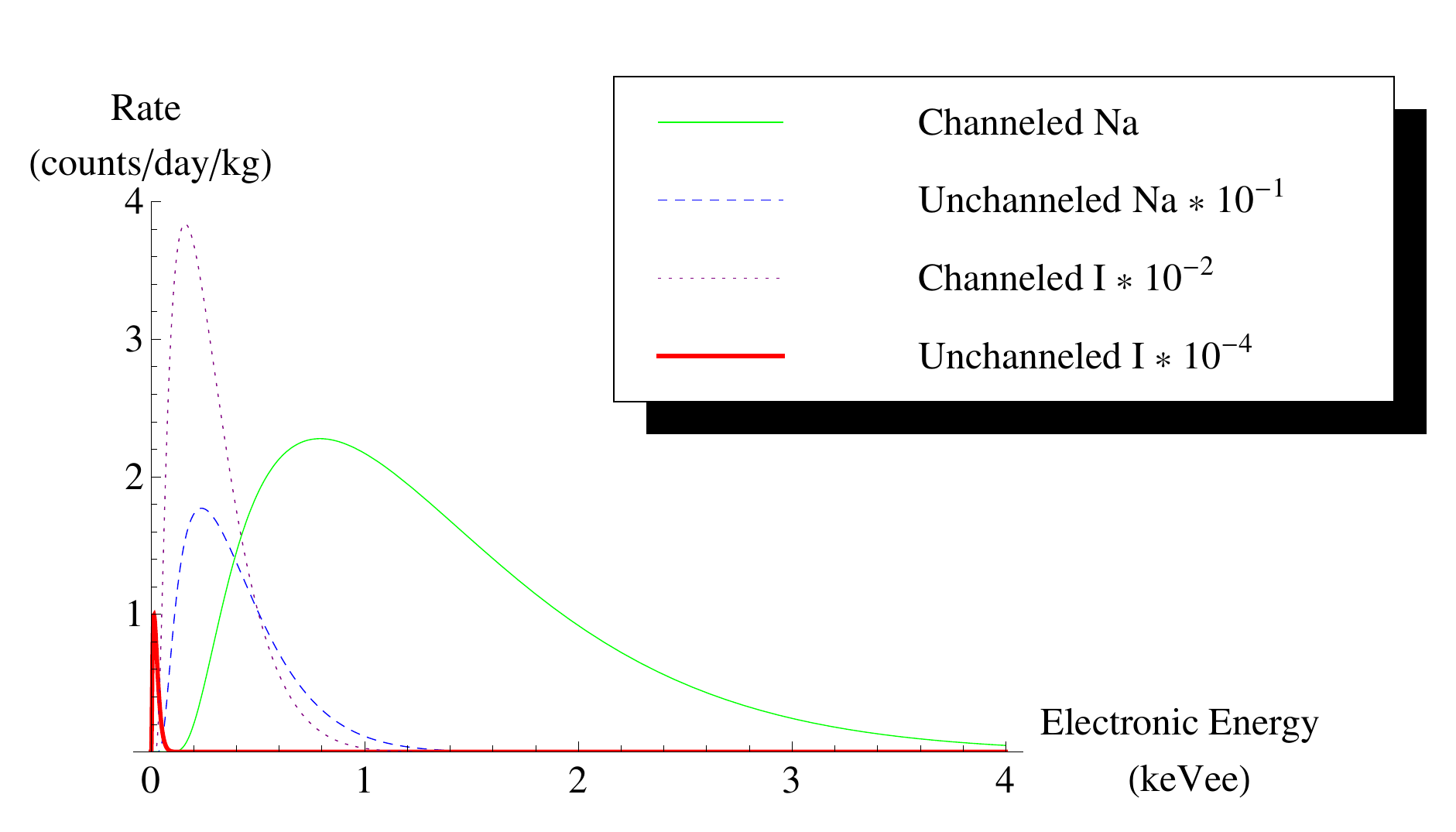}
\caption{ \label{Fig:DAMARatesConstant}   A sample of possible components of the unmodulated energy spectra at DAMA, arising from sodium and iodine scatters and channeled and unchanneled recoils. The spectra from iodine scatters are shown at 1\% of their actual scale. Here we have taken $\mchi = 3.5\GeV$ and $\delta = 6\keV$ and chosen the cross-section to fit the DAMA modulation signal. The channeling fraction has been assumed to be constant with energy and equal to 30\% for both sodium and iodine recoils. Note that only channeled sodium scatters are relevant for the energies where modulation data is available  ($> 2\keV$). Channeled sodium also dominates the rate at $2 \keV$ where the constraint form the DAMA unmodulated rate is strongest.}    
\end{center}
\end{figure}

The channeled and unchanneled rates for sodium and iodine scattering are shown (on different scales) in Figure \ref{Fig:DAMARatesConstant} for constant 30\% channeling. We fit the DAMA signal with channeled downscattering off of sodium in the NaI crystal. The rate of iodine scatters will be larger by a factor $\left( \frac{A_I}{A_{Na}} \right)^2 \approx 30$ but for our preferred parameter space will be almost entirely below threshold.  (The possibility of fitting the DAMA modulation using iodine scattering is discussed in section \ref{Sec:OtherOptions}).   

We calculate the rate at DAMA and other detectors in the standard way. For inelastic DM scattering through dimension-six operators
\begin{eqnarray}
\label{Eqn:Operators}
\frac{\bar{\chi}_1\chi_2\bar{q}q}{\Lambda^2} \hspace{.5 in} \text{or} \hspace{.5 in} \frac{(\phi_1\partial_\mu\phi_2-\phi_2\partial_\mu\phi_1)\bar{q} \gamma^\mu q}{\Lambda^2}
\end{eqnarray} 
with the dark matter either a fermion ($\chi_{1,2}$) or scalar ($\phi_{1,2}$), the single-proton cross section for downscattering at low momentum transfer is given by
\begin{eqnarray}
\label{Eqn:Sigman}
\sigma_n = \frac{C}{16\pi \Lambda^4}\left(\frac{\mchi m_n}{\mchi + m_n}\right)^2\sqrt{1+\frac{2\delta}{\mu_n v^2}}= \frac{C}{16\pi}\frac{\mu_n^2}{ \Lambda^4}\sqrt{1+\frac{2\delta}{\mu_n v^2}}
\end{eqnarray} 
where $\mu_n$ the reduced mass of the DM-nucleon system. Here $C = 4$ for fermions and 8 for scalars. We will take $C = 4$ in the remainder of the paper. Note that this cross-section goes as $\frac{1}{v}$ for splittings much larger than the kinetic energy, so the scattering rate $n \sigma v$ will be independent of velocity as stated above. In order to have a cross-section of $O(10^{-39} \text{ cm}^2)$ for a mass of a few GeV as preferred by the fit to DAMA we require $\Lambda \sim 340 \GeV$. Throughout this paper we will assume that the operators giving rise to elastic scattering are highly suppressed compared to the inelastic operators, which can be easily achieved in model-building \cite{IDMOriginal}. We will therefore ignore elastic scattering in the remainder of the paper.

It is useful to define the elastic limit of the above cross-section 
\begin{eqnarray}
\label{Eqn:Sigmanel}
\sigma_{n,el} = C \frac{\mu_n^2}{16\pi \Lambda^4}
\end{eqnarray} 
which is reached when the splitting is much less than the kinetic energy of the collision. The low-energy differential cross section is then
\begin{eqnarray}
\label{Eqn:DifferentialCrossSection}
\frac{d\sigma}{dE_R}=\frac{\sigma_{tot}}{\Delta E_R}= \frac{m_N}{2 \mu_n^2 v^2} \sigma_{n,el} [Z f_p + (A-Z) f_n]^2 F(q^2)
\end{eqnarray}
where $Z$ is the atomic number of target nucleus, $A$ the mass number, $f_p$ and $f_n$ constants that reflect the relative strengths of couplings to protons and neutrons, and $F(q^2)$ a form factor depending on the momentum transfer to the nucleus $q^2=2 m_N E_R$. We use the Helm form factor as given in \cite{Lewin&Smith}; because our signals peak at $q^2=2 m_N \times  \delta\frac{\mchi}{m_N}=2 \delta \mchi \sim (7 \MeV)^2$ this form factor is essentially negligible for scattering off of even very large nuclei.  When discussing the constraints on the model we will take $f_n = f_p$; however in Section \ref{Sec:Model} we discuss a model in which $f_n = 0$; i.e. the dark matter couples only to protons.
 
The differential scattering rate of dark matter per unit recoil energy $E_R$ is given by
\begin{eqnarray}
\label{Eqn:Rate}
\frac{dR}{dE_R}=N_Tn_\chi\int_{v_{min}(E_R)} \frac{d\sigma}{dE_R} v f(v) dv
\end{eqnarray}
\begin{equation}
v_{min} = \frac{|m_N E_R - \mu \delta|}{\mu \sqrt{2 m_N E_R}}
\end{equation}
where $N_T$ is the number of target nuclei, $n_\chi$ is the local number density of dark matter, $f(v)$ is the distribution of DM velocities relative to the target, and $v_{min}$ is the minimum velocity required to produce a recoil of energy $E_R$. $f(v)$ at a given time of the year is determined by the velocity of Earth through the halo and by the distribution of DM velocities within the halo itself, here assumed to be of the form
\begin{eqnarray}
\label{Eqn:HaloVelocity}
f_{Halo}(v_H)=\frac{N_0}{(\pi v_0^2)^\frac{3}{2}}e^{-\frac{v_{H}^2}{v_0^2}}\Theta(v_{esc}-v_{H})
\end{eqnarray} 
In the figures shown here we assume a Maxwell-Boltzmann distribution for the DM halo velocities with mean $v_0=220\kms$ and a sharp cutoff (i.e. galactic escape velocity) at $v_\text{esc} = 480 \kms$. A low escape velocity is preferable in our model as it increases the modulation fraction and reduces the tension with the DAMA unmodulated rate. $N_0$ is chosen to normalize the probability distribution to one.

Generically one expects the two mass states of dark matter to have equal relic densities. DM particles in the lower state can upscatter off of nuclei on Earth if their kinetic energies are greater than the splitting. The rate for upscattering is then determined by the same formulae as shown above but with $\delta \rightarrow -\delta$. For the parameter space favored by the fit to DAMA only the high-velocity tail of the dark matter distribtuion has sufficient energy to upscatter, so the rate is dominated by downscattering-- essentially only half of the dark matter density contributes to scattering on Earth. We will assume equal populations for the two mass states of dark matter throughout this paper.

The rate at DAMA at any given energy will be given by the sum of the channeled and unchanneled rates for both sodium and iodine scatters. These channeling fractions as a function of energy have not been measured in DAMA, but the collaboration has estimated them using a simple analytical model based on the work of Lindhard \cite{DAMAChanneling,Lindhard}. In this model the channeling fraction rises sharply as one goes to lower recoil energies, approaching 1 at zero energy; realistically one expects this increase to saturate at some point \cite{GelminiChanneling}.  In this paper we do not attempt to determine a realistic energy profile for the channeling fraction but simply assume constant 30\% channeling in both sodium and iodine unless otherwise noted. As discussed in section \ref{Sec:DAMAUnmod}, the constraint from the DAMA unmodulated rate is sensitive to the exact shape of the channeling fraction, but less so to the overall scaling. Because other dark matter search experiments are not sensitive to the DAMA preferred parameter space for exoDM, lower channeling fractions do not constrain the model, though for channeling fractions less than $\sim 10\%$ the fit to DAMA is undone by the contribution of the unchanneled sodium spectrum (for finite energy resolution). We have taken the quenching factors for unchanneled events to be constant and equal to .09 for iodine recoils and .3 for sodium.  

In fitting the DAMA modulation data we have modeled the energy resolution of the DAMA detectors using the relation given in \cite{DAMAApparatus}:
\begin{eqnarray}
\label{Eqn:EnergyResolution}
\frac{\sigma_E}{E} = \frac{\alpha}{\sqrt{\frac{E}{\keV}}}+\beta
\end{eqnarray}
with $\alpha = .448$ and $\beta = 9.1 \times 10^{-3}$.  

\begin{figure}
\begin{center}
\includegraphics[width = 7. in]{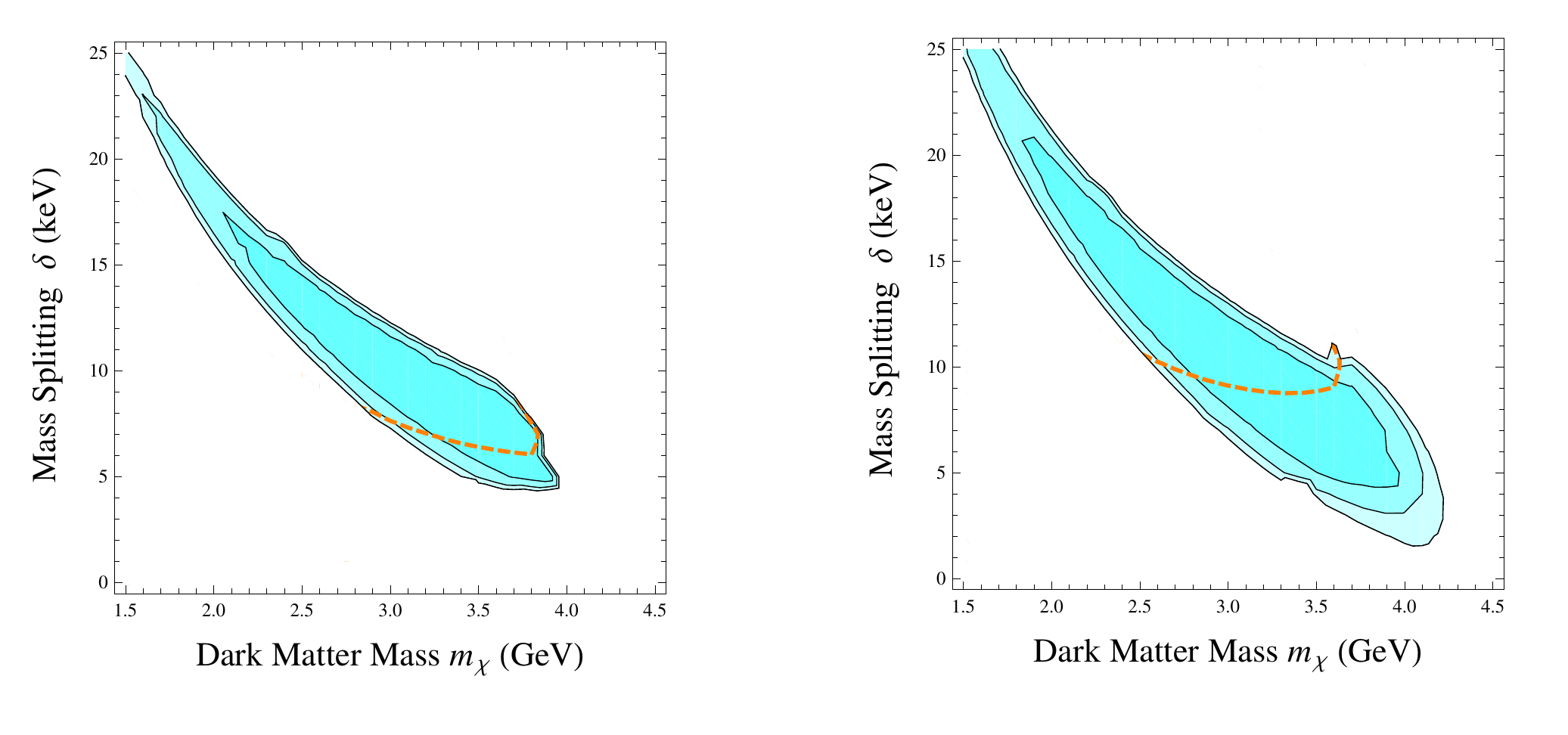}
\caption{ \label{Fig:ParameterSpaceConstant}  The parameter space in the mass/splitting plane for the downscattering fit to DAMA (68\%, 90\% and 99\% confidence level regions shaded). In the lefthand plot we assume constant channeling fractions for both sodium and iodine, while on the right we take the channeling fraction given by Eq. \ref{Eqn:DownturnChanneling}. At each point in the $(\mchi,\delta)$ plane in this plot the cross-section has been chosen to give the best fit to DAMA. The halo model is Maxwell-Boltzmann with $v_0 = 220 \kms$ and $v_{esc} = 480 \kms$. The orange dashed line indicates the constraints from the DAMA unmodulated rate at 2 keV (requiring less than 1 count/day/kg/keV-- see Section \ref{Sec:DAMAUnmod}). The null results from XENON10, CDMS-Si and other direct detection experiments do not further constrain this parameter space for the DAMA fit.}     
\end{center}
\end{figure}

The parameter space in the mass/splitting plane that fits the DAMA modulation signal under these assumptions is shown in the lefthand plot of Figure \ref{Fig:ParameterSpaceConstant}. At each point in this plot the cross-section has been chosen to optimize the fit to DAMA. The orange dashed line indicates the constraint from the unmodulated rate at 2 keV in DAMA. This constraint and the righthand plot are discussed in Section \ref{Sec:DAMAUnmod}.  

Although we fit the DAMA modulation with channeled sodium scatters, for sufficiently large dark matter masses the exponential tail of the spectrum of channeled iodine scatters appears in the modulation signal region (after smearing due to energy resolution) and rapidly erodes the goodness of fit. This accounts for the sharp cut-off of the DAMA-fit  parameter space as one increases the mass that can be seen in Figures \ref{Fig:ParameterSpaceConstant} and \ref{Fig:DownscatteringMSigma}. This behavior is of course highly dependent on the energy resolution and channeling fraction at recoil energies $\lesssim 1 \keVr$; in particular the effects of varying the channeling fraction can be seen in the righthand part of Figure \ref{Fig:ParameterSpaceConstant}.    

\section{Constraints}
\label{Sec:Constraints}

\begin{figure}
\begin{center}
\includegraphics[width = 7. in]{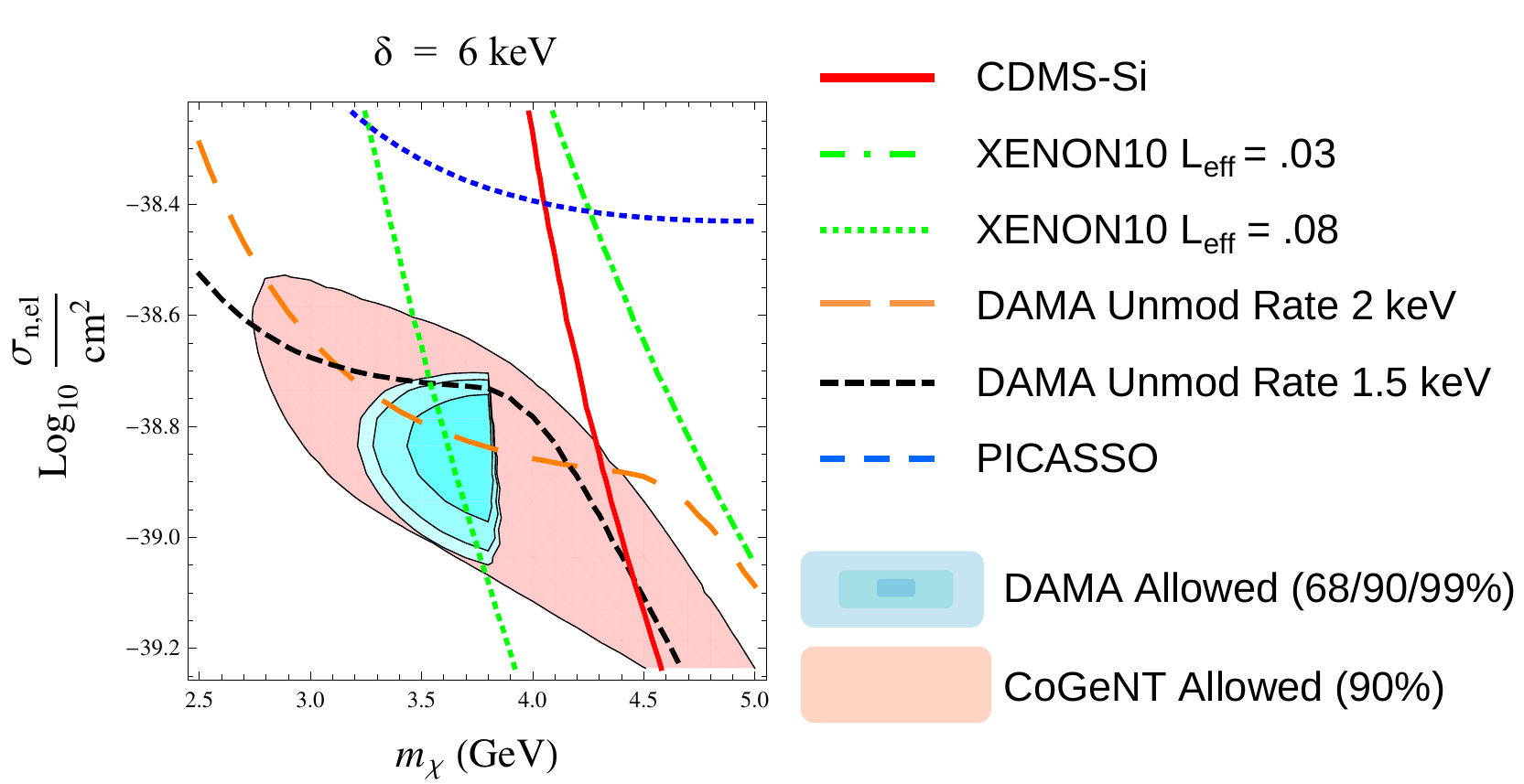}
\caption{ \label{Fig:DownscatteringMSigma}  The parameter space in the mass/cross-section plane for the exoDM fit to DAMA (68\%, 90\% and 99\% confidence level regions shaded in blue), with a mass splitting of $\delta=6\keV$ and assuming constant channeling fractions of 30\% for both sodium and iodine.  The halo model is Maxwell-Boltzmann with $v_0 = 220 \kms$ and $v_{esc} = 480 \kms$. The orange long-dashed, black short-dashed, red solid, green dot-dashed (dotted) and blue medium-dashed lines indicate respectively the constraints from the DAMA unmodulated rate at 2 keV, the DAMA unmodulated rate at 1.5 keV, the null results of CDMS (silicon), the null results at XENON10 for $\Leff =$ .03 (.08), and the null results from the PICASSO experiment (see Section \ref{Sec:Constraints}). The CDMS and XENON10 bounds are at 95\% confidence level, while the PICASSO curve is a conservative constraint in which we require the integrated rate from 2 to 5 keV to be less than 30 counts/day/kg. The red shaded region gives the parameter space to fit the CoGeNT excess at 90\% C.L. assuming a constant 5\% channeling fraction in germanium. The sharp cut-off of the DAMA fit region as the mass is increased is due to the leakage of the exponential tail of channeled iodine scatters into the modulation signal region as described in Section \ref{Sec:Kinematics}.}     
\end{center}
\end{figure}

\begin{figure}
\begin{center}
\includegraphics[width = 7. in]{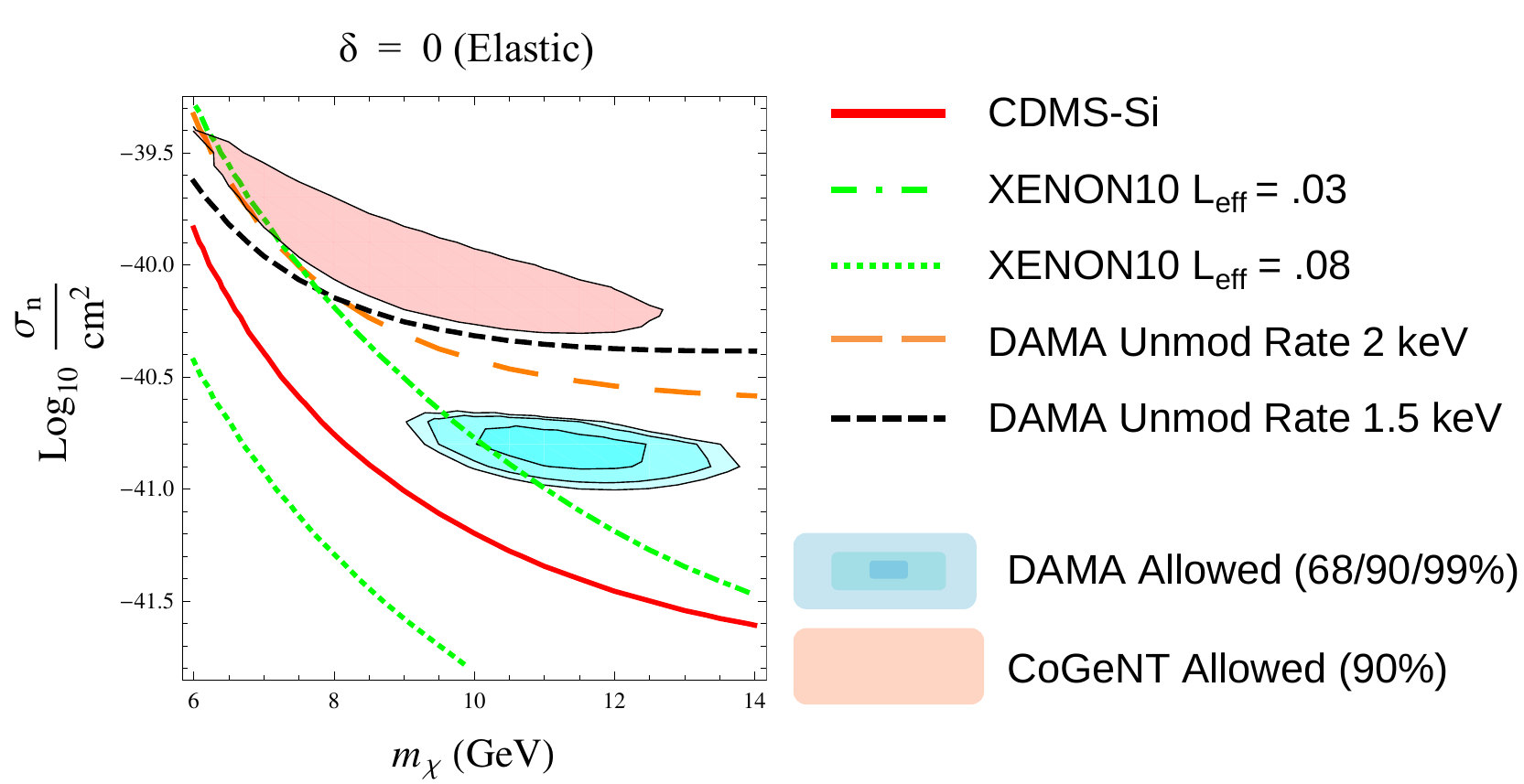}
\caption{ \label{Fig:ElasticMSigma}  The parameter space in the mass/cross-section plane for the elastic LDM (zero mass splitting) fit to DAMA (68\%, 90\% and 99\% confidence level regions shaded in blue), assuming constant channeling fractions of 30\% for both sodium and iodine. The halo model is Maxwell-Boltzmann with $v_0 = 220 \kms$ and $v_{esc} = 480 \kms$. The orange long-dashed, black short-dashed, red solid, and green dot-dashed (dotted) indicate respectively the constraints from the DAMA unmodulated rate at 2 keV, the DAMA unmodulated rate at 1.5 keV, the null results of CDMS (silicon), and the null results at XENON10 for $\Leff =$ .03 (.08) (see Section \ref{Sec:Constraints}). The CDMS and XENON10 bounds are at 95\% confidence level. The red shaded region gives the parameter space to fit the CoGeNT excess at 90\% C.L. assuming a constant 5\% channeling fraction in germanium. }     
\end{center}
\end{figure}

In this section we will consider the constraints on exoDM from other dark matter searches (including the DAMA unmodulated rate). In fact only the constraint from the DAMA unmodulated rate actually intersects the parameter space that fits the DAMA modulation signal, but this is highly sensitive to the details of the channeling fraction and efficiency of the DAMA detector at sub-threshold energies. The constraints from other experiments are shown in Figure \ref{Fig:DownscatteringMSigma} for exoDM; the constraints on elastically scattering light dark matter computed using the same methodology are shown in Figure \ref{Fig:ElasticMSigma}. In particular we note that under the assumptions of our analysis the elastic LDM explanation for DAMA \cite{Fairbairn:2008gz,Chang:2008xa,ModerateChanneling,Fitzpatrick:2010em} is completely ruled out. In the following sections we describe each of the constraint curves in these plots.  

\subsection{Constraints from DAMA Unmodulated Rate}
\label{Sec:DAMAUnmod}

As illustrated in Figure \ref{Fig:DAMARatesConstant}, the modulating rate is a very small fraction of the total scattering rate at DAMA in this model. It is therefore important to check that the model is also consistent with the unmodulated rates observed at DAMA. Although the DAMA collaboration considers the energy threshold of their experiment to be 2 keVee and have not released data on the modulation signal below those energies, they have published the unmodulated rates for their earlier runs down to electronic energies of $\sim 1 \keV$ \cite{DAMAFirstResults}. As bounds on the expected background at these energies are not available, we will constrain the model by requiring that the rate of DM scatters at a given recoil energy not exceed the total unmodulated rate observed at that energy. In particular the rate at $\sim 2 \keVee$ gives the strongest constraint. (If one assumes that the background at this energy is equal to the observed background at energies $6 - 10 \keV$, then both exoDM and elastically scattering light dark matter are essentially ruled out \cite{Chang:2008xa}.) The unmodulated rate at these energies is dominated by channeled sodium scatters in our model (as shown in Figure \ref{Fig:DAMARatesConstant}); for most of the parameter space that fits the DAMA modulation signal this rate is greater than the observed $\sim 1 \text{ count/day/kg/keV}$ (Figure \ref{Fig:ParameterSpaceConstant}). The rate at  $\sim 1.5 \keVee$ can also be somewhat constraining as shown in Figure \ref{Fig:DownscatteringMSigma}. 

We note that there are several possibilities that might relax these constraints. In particular, we emphasize the unknown behavior of the channeling fractions and quenching factors at low energies. If the channeling fraction is significantly larger at energies below 2 keV than it is in the modulation signal region, the tension with the DAMA unmodulated rate increases; conversely if the channeling fraction is lower at these energies then the unmodulated rate is less constraining. In the model of \cite{DAMAChanneling} the channeling fraction increases with energy down to arbitrarily low energies. Such behavior for the channeling fraction would cause exothermic dark matter to produce too high an unmodulated rate at energies $< 2 \keV$, ruling out the parameter space that fits the modulated signal. In reality however one should not expect this simple Lindhard model of channeling to hold for arbitrarily large channeling fractions; in particular due to geometrical blocking in the crystal the channeling fraction may saturate or decrease as one goes to sufficiently low energies \cite{GelminiChanneling}. Such uncertainties in the channeling fraction of course affect the exact parameter space that fits DAMA. 

To illustrate the possible effect of a drop in the channeling fraction at low energies we show in the righthand plot of Figure \ref{Fig:ParameterSpaceConstant} the parameter space to fit DAMA assuming a channeling fraction of the form
\begin{eqnarray}
\label{Eqn:DownturnChanneling}
f_{channel} =
\left\{ 
\begin{array}{l l}
  .3 & \quad E_R \geq 3 \keV\\
  .3 \times \frac{E_R}{3 \keV} & \quad E_R < 3 \keV\\
\end{array} \right.
\end{eqnarray}
We see that this ``model" for the channeling fraction, in which the channeling at 2 keV is two-thirds of that at higher energies, considerably increases the allowed parameter space. 

Uncertainties in the detector efficiency close to threshold can also be important for determining the constraint from the DAMA unmodulated rate. DAMA has published samples of efficiency measurements down to 2.5 keV \cite{DAMAApparatus}; if the efficiencies at lower energies (below threshold) are systematically lower than the values used in determining the unmodulated rates then the above constraints may be relaxed. Uncertainties in the energy calibration and resolution below threshold may also be relevant. 

We note that the rate at $\sim 1 \keVee$ can be affected by upward fluctuations from the channeled iodine spectrum due to finite energy resolution. The rate in the 1 keVee bin then depends sensitively on the paramterization of the energy resolution, in particular whether equation \ref{Eqn:EnergyResolution} is used or if the upward fluctuations are described by Poissonian statistics based on the measured values of 5-7 photoelectrons/keV \cite{DAMAApparatus}. Because of this uncertainty in addition to that of the sub-keV channeling fractions and detector efficiency we do not consider the rate at 1 keVee as constraining the model.  

\subsection{Constraints from XENON10}
\label{Sec:XENON10}

The XENON10 experiment has a relatively low threshold of $\sim 2 \keV$ and may also be sensitive to exoDM. Although the expected rate of nuclear recoils with energy $> 2 \keV$ at XENON10 is essentially zero, lower energy events may be intepreted as $\sim 2 \keVr$ events due to upward fluctuations in the photoelectron count from such events. In particular, in order for an event to pass various cuts at XENON10 a sufficient number of photoelectrons must be observed in both the S1 (scintillation) and S2 (ionization) pulses. Assuming Poissonian statistics for the photoelectron and ionization electron counts one can determine the probability for a low energy event to be detected in XENON10. In deriving the constraints in this paper we require that an event produce three S1 photoelectrons and the equivalent of 12 ionization electrons in the S2 pulse in order to be identified as a WIMP scatter candidate, based on the acceptance and cuts described in \cite{Sorensen}. Because the spectrum from downscattering in xenon drops off sharply above 1 keV, the XENON10 constraints are highly dependent on the detector response at sub-keV energies where reliable measurements do not exist. We use as a guide the measurements of scintillation efficiency and ionization yield displayed in \cite{Manzur}, which show a downward trend in scintillation efficiency $\Leff$ at low energies (unlike the results of \cite{Sorensen}). In the absence of data at energies below 4 keV we will simply parameterize $\Leff$ as constant and consider two possible values: $\Leff = .08$ and $\Leff = .03$. $\Leff = .08$ is a conservative choice if we are guided by the measurements of \cite{Manzur}, as it implies no drop in $\Leff$ below 4 keV. $\Leff = .03$ is approximately the one-sigma lower bound at 4 keV from \cite{Manzur}. We take the ionization yield to be constant in the regime $0 - 4 \keVr$ and equal to the value measured by \cite{Sorensen} at 2 keVr, 12 photoelectrons/keVr. We consider the XENON10 dataset of \cite{Xe10Results}, a 58.6 day exposure for which no events were reported below 15 keV. 

The 95\% confidence level constraints from XENON10 for the two choices of $\Leff$ are shown in Figure~\ref{Fig:DownscatteringMSigma} for the downscattering fit to DAMA and in Figure~\ref{Fig:ElasticMSigma} for elastically scattering light dark matter. Our results imply that XENON10 only begins to constrain the model for scintillation efficiencies of $\Leff \gtrsim .08$ for sub-keV recoil energies. We note that even for $\Leff = .03$ (the one-sigma lower value) the elastic light dark matter scenario remains inconsistent with XENON10 once the abovementioned upward fluctuations in energy measurements are accounted for. We note that the assumptions about $\Leff$ we have taken lead to bounds that are more stringent than those of \cite{Fairbairn:2008gz,ModerateChanneling,Chang:2008xa,Fitzpatrick:2010em,Interpretations}, which to our knowledge do not account for Poissonian fluctuations in the scintillation photon counts. Again, such bounds depend strongly on the behavior of $\Leff$ at low energies where reliable measurements do not exist; we have shown here that the possible constraints on the exoDM parameter space are less stringent than those for elastic light dark matter.

We note that the ZEPLIN-III experiment \cite{Lebedenko:2008gb,Akimov:2010vk}, another liquid xenon detector with a large exposure and low threshold, may also be relevant for constraining exoDM (and elastic light dark matter). The ZEPLIN collaboration has measured  $\Leff$ under their operating conditions down to energies of $\sim 7 \keV$ \cite{Lebedenko:2008gb}; at the lowest point $\Leff \lesssim .03$. With this scintillation efficiency at low energies the bounds on exoDM are not constraining. However if the scintillation efficiency is closer to the values measured by \cite{Manzur} then the constraints likely become more significant; again the low-energy response and acceptance of the detector is critical in determining the expected rate from exoDM.         

\subsection{Constraints from CDMS Silicon}
\label{Sec:CDMS}

The silicon detectors of the CDMS experiment can be sensitive to light dark matter due to their low atomic mass and low threshold energy (5 keV for the analysis of \cite{CDMS-SUF} compared to 10 keV for the germanium analyses). We use the efficiency function given in \cite{Savage:2008er} in computing the expected rates in the CDMS silicon detectors. We include data from the runs described in \cite{CDMS-SUF} (5 keV analysis threshold) and \cite{CDMSTwoTower,FilippiniThesis} (7 keV threshold). No WIMP-scatter candidate events were identified at energies $< 50 \keV$ in the silicon detectors for these runs. We consider the rates at energies up to 20 keV in setting bounds on light dark matter as the DM signal is negligible at higher energies. In computing limits we assume an expected count rate of 1.2 from backgrounds in all runs combined, and take the 95\% confidence level bound to correspond to 3 expected events total. The constraint from CDMS is shown in Figure~\ref{Fig:DownscatteringMSigma}; this constraint does not intersect the parameter space preferred by the light exoDM fit to DAMA, while strongly constraining the elastically scattering light dark matter scenario. The differences between our results and those of \cite{Fairbairn:2008gz,ModerateChanneling,Chang:2008xa,Fitzpatrick:2010em,Interpretations} are mainly due to our inclusion of multiple runs, in particular the 5 keV threshold CDMS-SUF run \cite{CDMS-SUF} which we find places the strongest constraints. As discussed in \cite{Interpretations}, adjusting the halo velocity distribution does not significantly reduce the tension between CDMS Silicon and the DAMA fit region in the elastic scattering case.

\subsection{Possible Signal at CoGeNT}
\label{Sec:CoGeNTSignal}

\begin{figure}
\begin{center}
\includegraphics[width = 3.7 in]{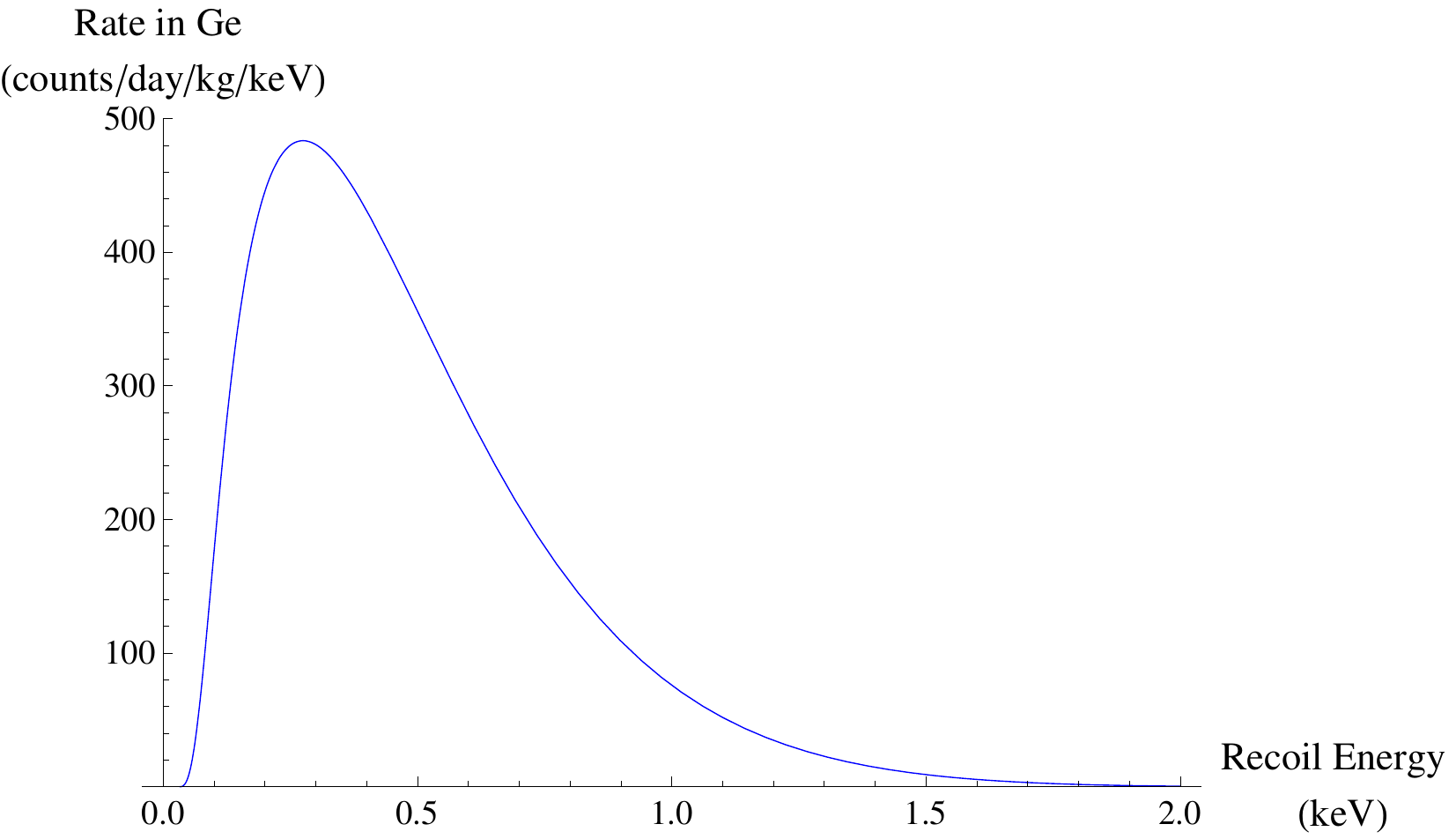}
\caption{ \label{Fig:GeRate}  The spectrum of recoil energies for downscattering off of germanium with parameters that fit the DAMA modulation:  $\mchi = 3.5\GeV$, $\delta = 6\keV$ and $\sigma_{p,el} = 1.4 \times 10^{-39} \text{cm}^2$. The halo model is Maxwell-Boltzmann with $v_0 = 220 \kms$ and $v_{esc} = 480 \kms$. }     
\end{center}
\end{figure}

Recently the CoGeNT experiment has reported an excess of events at low energies ($.5 - 1 \keVee$) that may be a signal of light dark matter  \cite{CoGeNT}. Assuming standard quenching this energy range corresponds to approximately $2 - 4 \keVr$. As shown in Figure \ref{Fig:GeRate}, the rate of germanium scatters predicted by our downscattering model is negligible in this energy range. However, if some fraction of nuclear recoil events are channeled in germanium then downscattering can fit the signals at both DAMA and CoGeNT. As shown in Figure \ref{Fig:DownscatteringMSigma}, the CoGeNT excess and the DAMA modulation can be fit simulataneously assuming a $\sim 5\%$ channeling fraction in germanium in the range $.5 - 1 \keVr$. (This is assuming constant $30 \%$ channeling at DAMA; for other sodium channeling fractions the germanium channeling fraction must be scaled accordingly.) The channeling fraction in germanium at these energies is unknown, but can potentially be measured or constrained with calibration data in germanium experiments. We fit the CoGeNT data with the channeled exoDM signal plus a constant background plus two gaussian peaks of fixed spread and relative height located at 1.1 and 1.3 keVee (corresponding to the expected lines from $^{68}$Ge and $^{65}$Zn). We do not inlcude an exponential background in the fit as in \cite{CoGeNT,Fitzpatrick:2010em}. Note that with such a germanium channeling fraction the predicted rate above the CDMS germanium threshold of 2.7 keVee is still consistent with the observed null results.

\subsection{Constraints from other experiments}
\label{Sec:OtherExp}

\subsubsection{PICASSO}
\label{Sec:PICASSO}

The PICASSO experiment \cite{Archambault:2009sm} operates in a manner similar to a bubble chamber: it uses superheated droplets of \ce{C4F10} and measures the bubble formation rate as a function of temperature. Although the energy of bubble events cannot be measured, the energy threshold for a recoiling nucleus to produce a bubble varies with temperature. By scanning the temperature, the integrated rate below various energies can be observed. The energy threshold can be adjusted to as low as $\sim 2 \keV$. This low threshold combined with the light target nuclei give PICASSO sensitivity to light dark matter. In Figure \ref{Fig:DownscatteringMSigma} a constraint curve for PICASSO is shown corresponding to a rate of 30 counts/kg/day integrated from 2 to 5 keV (assuming identical detector response to carbon and fluorine recoils). This bound is rather conservative given the data of \cite{Archambault:2009sm}. However, with a reduction of the background rate and/or measurement uncertainties PICASSO could in principle probe the parameter space of spin-independent exoDM. Exothermic scenarios involving spin-dependent scattering off of protons are highly constrained by PICASSO due to the high spin content of fluorine, as described in section \ref{Sec:OtherOptions}.  

\subsubsection{KIMS}
\label{Sec:KIMS}

The KIMS experiment \cite{KIMS} measures electron equivalent energies of events in CsI crystals and uses pulse shape discrimination to identify nuclear recoil events. The energy threshold of the experiment is 3 keVee. Because both of the elements of the CsI crystal are heavy, exoDM produces no signal above this threshold. Depending on the channeling fraction in CsI at $\sim 3 \keV$ the KIMS results could be a relevant constraint for elastically scattering light dark matter, however.

\subsubsection{CRESST-I}
\label{Sec:CRESST-I}

The CRESST-I experiment \cite{CRESST-I} uses detectors of sapphire (\ce{Al2O3}) crystal. The low energy threshold and the presence of a light element would favor this detector technology as a probe of light dark matter and exoDM in particular, however the large background at energies below $\sim 7 \keV$ prevents this experiment from placing a competitive constraint. 

\subsubsection{COUPP}
\label{Sec:COUPP}

The COUPP experiment \cite{COUPP} also uses a bubble-chamber-like operating principle, using superheated \ce{CF3I}. The pressure is varied to adjust the bubble nucleation threshold. An initial data set with backgrounds from alpha decays does down to energy thresholds of 5 keV. Preliminary results from a first run includes data with a 20 keV threshold~\cite{COUPPTalk}.
As with PICASSO, the presence of light elements would allow this experiment to probe exoDM if the energy threshold can be lowered somewhat. We discuss the expected spectrum in the next section.

\subsubsection{SIMPLE}
\label{Sec:SIMPLE}

The SIMPLE experiment \cite{SIMPLE} also uses superheated droplet technology, in this case using superheated \ce{C2ClF5}. Though their background is significantly lower than PICASSO or COUPP. the expected rate from exoDM above the threshold of 8 keV is exponentially small; again lower thresholds are required to probe the preferred parameter space.

\subsubsection{TEXONO}
\label{Sec:TEXONO}

The TEXONO experiment uses a Ge crystal detector with a very low threshold energy of 200 eV \cite{Lin:2007ka}.  Given their relatively small number of kg-keV-days this experiment does not provide a stringent constraint on exoDM parameter space.  Further, the techniques used by the collaboration to determine the signal region and acceptance for the dark matter search results has been called into question \cite{Avignone:2008xc} and it is unclear how reliable the actual limit is.  Thus, TEXONO does not provide a limit on exoDM, though if these issues were resolved and the threshold could be reliably kept so low such an experiment with larger exposure could potentially provide a good test of exoDM.

\section{Predictions}
\label{Sec:Predictions}
Nuclear recoils from exothermic dark matter interactions peak at energies $\sim \delta \frac{m_\chi}{m_N}$. These interactions lead to higher energy recoils in light nuclei. Additionally, owing to constraints from a variety of other experiments (see section \ref{Sec:OtherOptions}) the DAMA signal is most easily fit with sodium recoils in the energy range $\sim 2 - 3$ keV. Since sodium is a light nucleus, this model will give rise to signals in the few keV band in experiments that employ light nuclei while the recoils will generically be in the sub keV band for heavy nuclei. Experiments that have a low threshold energy with light nuclei are therefore the best options for discovering these exothermic dark matter interactions. 

\begin{figure}
\begin{center}
\includegraphics[width = 4.1 in]{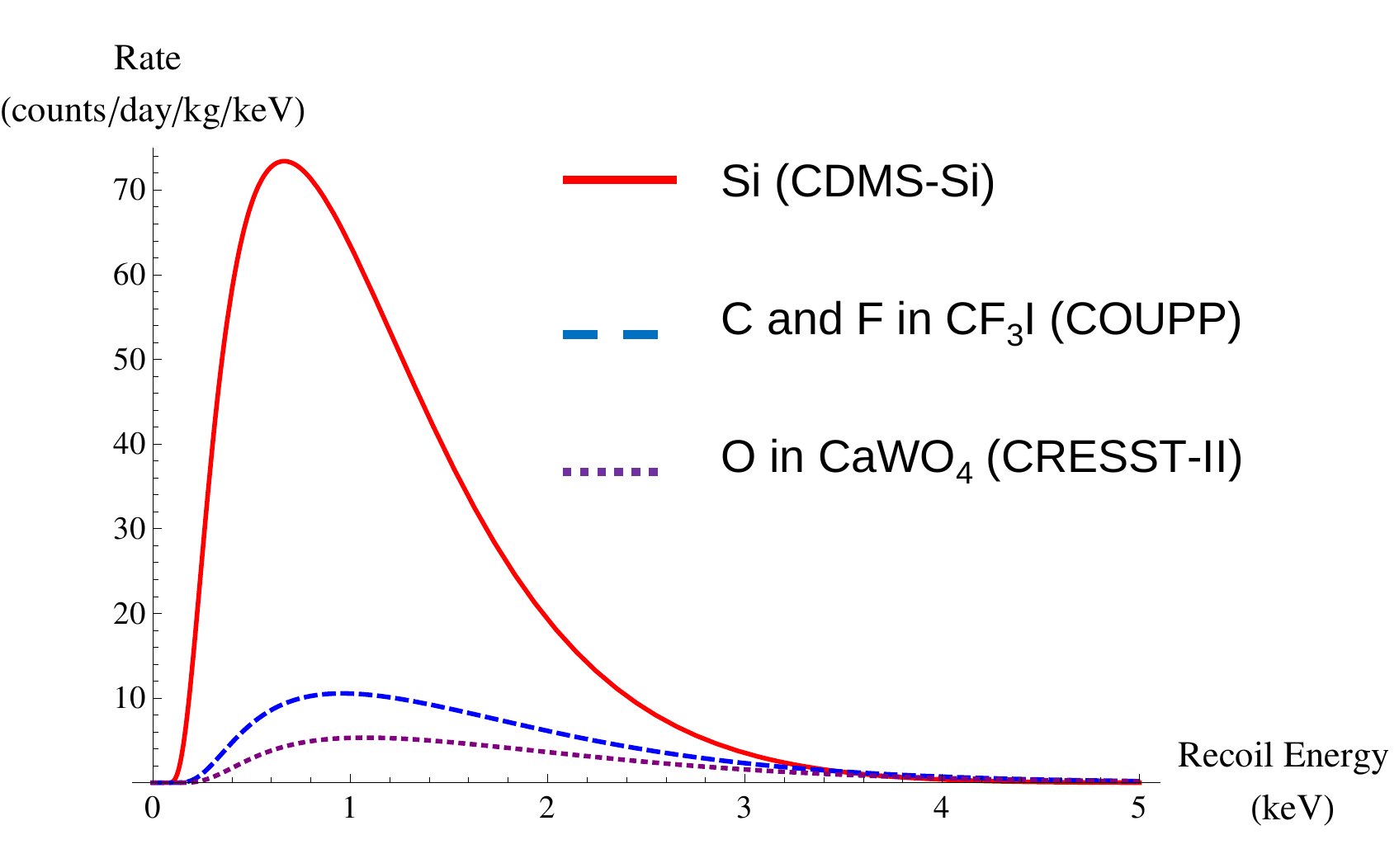}
\caption{ \label{Fig:LowEnergyRates}  The spectra for recoils off of light nuclei in the detector materials of CDMS-Si, COUPP, and CRESST-II. Detector response effects (such as the efficiency in CDMS-Si, which is ~5\% at 5 keV) have not been included. The parameters are the same as in Figure \ref{Fig:GeRate}.}     
\end{center}
\end{figure}

For example, the parameter space that fits the DAMA region gives rise to a large number of events below $\lessapprox 4$ keV in CDMS Silicon (see Figure \ref{Fig:LowEnergyRates}), while  the rate is exponentially suppressed above $\sim 5$ keV. Similarly, a large number of events are also expected in CRESST and COUPP in this region since they also contain light nuclei (see Figure \ref{Fig:LowEnergyRates}). Currently, the thresholds of these experiments are above $\gtrapprox 5$ keV. If these thresholds can be reduced to $\sim 1 - 2$ keV, then even in the presence of significant backgrounds these experiments can serve as a stringent test of this framework. 

While experiments with light nuclei are the best vehicles to discover these models, experiments with heavy nuclei can also place constraints on this framework. This is because the scattering cross-section scales as $\sim A^{2}$ giving rise to a significantly larger scattering rate off of heavier nuclei, albeit at lower energies. An improved understanding of the low energy response of these detectors can be used to constrain these models. As discussed in section \ref{Sec:DAMAUnmod}, the most severe constraint on the exoDM fit to the DAMA signal arises from the unmodulated DAMA spectrum below 2 keV. A better understanding of the detector, both in terms of the channeling fraction and the detector efficiency at these energies, will further constrain exoDM.  Similarly, as discussed in section \ref{Sec:XENON10}, measurements of the scintillation efficiency of xenon in the sub-keV region can also restrict these models.  Additionally,  CDMS Germanium may also be able to probe this framework if its energy threshold is pushed to around $\sim 1$ keV. In this case, the tail of the exothermic spectrum should become visible (see Figure \ref{Fig:GeRate}).

\section{Other Options}
\label{Sec:OtherOptions}
The above discussions were centered around the possibility of explaining the observed annual modulation in DAMA with light ($\lesssim 5$ GeV)  exothermic dark matter with splittings $\delta \sim 5 - 10$ keV. In this parameter space, the sodium nuclei at DAMA are kicked by $\sim 2-3$ keV while the iodine recoils occur at $\ll 1$ keV. A fraction of these events are channeled, leading to observable scattering off sodium nuclei in DAMA. By varying the mass splitting $\delta$, the energy that is deposited in the recoiling nucleus is altered. As $\delta$ increases, there is a proportional rise in the deposited energy. This gives rise to other possible ways to explain the DAMA signal, in particular through unchanneled sodium recoils as well as channeled and unchanneled iodine recoils. In this section, we briefly examine and constrain these possibilities. 

\subsection{Unchanneled Iodine} 
\label{Sec:UnchanneledIodine}
The quenching factor for scattering off of iodine in NaI is $\sim 0.09$ \cite{Lewin&Smith}, giving rise to a signal region $\sim 22 - 30$ keV. An exothermic, spin-independent dark matter - nucleus interaction in this region leads to events in XENON10 in the same energy band and is severely constrained \cite{Xe10Results}. In particular, \cite{Xe10Results} constrains the modulation fraction in DAMA to be over 21$\%$. As discussed earlier,  exothermic dark matter typically leads to modulation fractions $\lessapprox 1\%$. Consequently, it is difficult to reconcile the DAMA data with the observations of XENON10 for exothermic, spin-independent interactions. However, somewhat larger modulation fractions may be realized with non-standard dark matter velocity profiles. We emphasize that the dark matter velocity distribution has not been experimentally measured and there are significant differences amongst various theoretical expectations of this distribution. In particular, the addition of a dark disk could potentially enhance the modulation fraction \cite{Bruch:2008rx}. 

Spin-dependent interactions are also constrained by XENON10. Roughly $\sim 50 \%$ of naturally occuring xenon (which is used by the XENON10 collaboration) carries a nuclear spin, leading to spin-dependent interactions with the dark matter. A naive extrapolation of  \cite{Xe10Results} would then suggest that the conflict between XENON10 and DAMA can be resolved if the modulation fraction is $\approxeq 10\%$, which is also difficult to achieve for exothermic dark matter. However, the nuclear spin in xenon is carried by the neutron while the nuclear spin in DAMA (for either sodium or iodine) is carried by the proton. Consequently, a spin-dependent interaction that couples more strongly (by a factor of $\sim 10$) to protons than neutrons would lead to consistency between DAMA and XENON10.

Stringent bounds on spin-dependent proton-dark matter interactions are placed by PICASSO \cite{Archambault:2009sm}. The target mass in PICASSO is C$_{4}$F$_{10}$ with the nuclear spin being carried by the proton in fluorine. The DAMA region leads to an unacceptably large number of fluorine recoils around $\sim 150$ keV in PICASSO.  This  is due to both the large number of fluorine atoms in C$_{4}$F$_{10}$ and the enhancement in the nuclear spin interactions of fluorine \cite{Lewin&Smith}. The absence of such a high rate in the data of \cite{Archambault:2009sm} rules out this possibility.

\subsection{Channeled Iodine} 
\label{Sec:ChanneledIodine}
Channeled iodine events can explain the DAMA signal if the exothermic interactions dump energies $\sim 2 - 3$ keV in the iodine nuclei. With a low modulation fraction $\lessapprox 1 \%$, this will also lead to some events at $\sim 9 - 12$ keV at CDMS Silicon for spin-independent interactions. The null results at CDMS Silicon strongly constrain this possibility. However, with larger modulation fractions $\gtrapprox 5 \%$, there are regions of parameter space where the DAMA observations can be reconciled with the absence of events at CDMS Silicon. These larger modulation fractions are difficult to incorporate into exothermic dark matter models. But, as discussed earlier, these modulation fractions could perhaps be attained with non-standard dark matter velocity profiles. Furthermore, the modulation fraction can be increased if the scattering operators are momentum dependent. 

We also note that the light, spin-independent, elastic dark matter scenario \cite{Fitzpatrick:2010em} which explains DAMA through channeled iodine scattering is significantly constrained by CDMS Silicon. In particular, the inclusion of all  CDMS Silicon data sets essentially rules out this parameter space \cite{Chang:2008xa} (see Figure \ref{Fig:ElasticMSigma}). Recall that downscattering leads to peaks at higher energies for lighter elements, so as one increases the splitting one increases the energies of events at CDMS-Si more than the energies of iodine events in DAMA. Therefore we find that the fit to DAMA for channeled downscattering off of iodine is even more constrained by CDMS-Si than the elastic scattering scenario. 

Because silicon does not contain an unpaired proton, one might imagine that spin-dependent couplings to the proton would allow exoDM to bypass the CDMS constraint. The most stringent constraints on spin-dependent interactions are placed by PICASSO. In this scenario, the DAMA events would imply events at PICASSO  (from fluorine scattering) at energies $\sim 14 - 20$ keV. As explained in section \ref{Sec:UnchanneledIodine}, the large number of fluorine atoms in C$_4$F$_{10}$ and the enhancement in the nuclear spin interactions of fluorine gives rise to an unacceptably large rate in PICASSO \cite{Archambault:2009sm}. Consequently, this  spin-dependent scenario is  ruled out.

\subsection{Unchanneled Sodium}
\label{Sec:UnchannelledSodium}
In the absence of channeling, the DAMA signal region can be reproduced by sodium scattering events if these events are at energies $\sim 8 - 9$ keV. Even with nearly 100\% modulation, this leads to events at $\sim 6.5 - 7.5$ keV in CDMS Silicon for spin-independent interactions. The absence of such events rules out this possibility. Similarly, spin-dependent interactions will lead to events at PICASSO (from fluorine scattering) around $\sim 10$ keV. The absence of such events rules out this scenario. 

\section{A Theory of Exothermic dark matter}
\label{Sec:Model}

Thus far we have worked within an effective theory of exoDM in which two nearly degenerate states are added which interact with the standard model only via one  of the operators of equation~(\ref{Eqn:Operators}). A more complete model may explain the small mass splitting dynamically and also explain why operators that lead to elastic scattering are suppressed or absent; this has been explored in the context of upscattering inelastic dark matter \cite{IDMOriginal, Cui:2009xq, Bagnasco:1993st, Alves:2009nf}.

One possibility is to introduce a light Abelian gauge boson $A'$ (a ``dark photon'') of mass $m_{A'}$ that mixes kinetically with the Standard Model $U(1)$ and couples to the ``dark sector'' as in~\cite{AToDM}.  The strength of the gauge coupling in the dark sector is set by a dark fine structure constant $\alpha_D$ and the kinetic mixing parameter is defined by  $\frac{\epsilon}{2} F_{\mu \nu}' F^{\mu \nu}$. This package of model building was proposed for dark matter in the context of inelastic up-scattering and recent cosmic ray anomalies. It can explain the inelasticity of dark matter couplings and the small mass splitting in the dark sector, of order $\alpha_D m_{A'}$. In the original models the mass scales of dark matter were taken to be at a TeV with small MeV splittings but we will adapt this framework to smaller scales and splittings.

For the purpose of theories of exoDM , we will take the scale of masses in the dark sector to be lighter, of order few GeV, and the splitting to be of order a few to ten keV. These two scales are motivated by the kinematics of exoDM which as discussed in the previous sections can explain the DAMA and CoGENT signals while evading other bounds. The mass splitting of order $\alpha_D m_{A'}\sim 10$ keV is naturally generated in this framework for a dark photon mass of $\sim 100$ MeV and a dark fine structure constant of $\alpha_D\sim 10^{-4}$. Interestingly this gives the correct rate in DAMA and CoGENT. Even with higher dark photon masses the correct spectrum may arise with additional model building or tuning. 

In the low energy theory below ($ m_{A'}$) we generate the operators in equation~(\ref{Eqn:Operators}) with
\begin{eqnarray}
\label{Eqn:KineticLambda}
\Lambda=\frac{m_{A'}}{(16\pi^2 \alpha_D \epsilon^2 \alpha)^{1/4}}
\end{eqnarray}   
where $\alpha_D$ gives the coupling of the new gauge boson to the dark matter. 
It is very simple to get the correct scale of $\Lambda\sim 340$ GeV, for example with 
\begin{equation}
m_{A'}=  \left(\frac{\epsilon}{10^{-5}}\right)^{1/2} \times \left(\frac{\alpha_D}{10^{-4}}\right)^{1/4}\times
100\mbox{ MeV}
\end{equation}

Constraints on light $U(1)$'s kinetically mixed with the Standard Model exist from beam dump experiments, $e^+e^-$ colliders and lepton anomalous magnetic moments and have been compiled in \cite{Bjorken:2009mm}.  Searches for $A'$s is decays of the $\Upsilon$(3S) places a bound for $\epsilon\lesssim 0.003$. However, for $m_{A'}$ above twice the dark matter mass the dark photon will primarily decay invisibly to dark matter pairs. The bound then is $\epsilon\lesssim 0.05$ from semi-invisible $\Upsilon$ decays~\cite{babar}.
However a large portion of our parameter space is yet to be probed, but may be explored in upcoming fixed target experiments~\cite{Bjorken:2009mm,apesex}.

Within the context of a kinetic mixing model we will now discuss the relic abundance of dark matter and the fraction of dark matter in the excited state today.
Begining with the overall abundance, for $\mAp < \mchi$ the dark matter will predominantly annihilate to two dark photons, which will decay to SM particles. The cross-section for this process is $\sim \frac{\alpha_D^2}{\mchi^2}$; we can achieve a thermal relic abundance for the dark matter by taking this to be $\sigma_0=10^{-9} \GeV^{-2}$. For example we can get this by taking $\alpha_D \sim 10^{-4}$ for $\mchi = 3 \GeV$. For larger $\alpha_D$ the thermal abundance will be smaller than the observed dark matter density and out-of-equilibrium processes are necessary to produce the observed dark matter density.

The exothermic dark matter scenario requires that a relic abundance of the excited dark matter state exist today. This requires that (a) scattering processes that cause deexcitation freeze out at temperatures greater than the mass splitting between the two states and (b) the excited state has a lifetime longer than the age of the Universe. Two excited state dark matter particles can go to two ground state particles through dark photon exchange with cross section $\sigma \sim \alpha_D^2 \frac{\mchi^2}{\mAp^4}$. For the mass and splittings preferred by the fit to DAMA and taking $\alpha_D = 10^{-4}$, we obtain a bound on $\mAp \apprge 100 \MeV$ in order for deexcitation to freeze out sufficiently early. 

As for the lifetime of the excited state, the only possible decay channels for the excited state are to neutrinos or three photons; these have been explored in the context of kinetic mixing with the Standard Model in \cite{MetastableWIMPs, pospelov}. These decays are highly suppressed due to the smallness of the splitting, so the excited state lifetime is generically much longer than the age of the Universe. If fact, additional model-building is required to obtain a conventional inelastic (up-scattering) dark matter scenario where only the lowest mass state exists today~\cite{MetastableWIMPs, pospelov}. 

\section{Indirect Constraints}
\label{Sec:Indirect}

\subsection{Annihilation in the Sun}

Dark matter particles with sufficiently large nuclear interaction cross sections are expected to be captured by the Sun and annihilate in its interior. These annihilations may produce high energy neutrinos that escape the Sun.  The Super-Kamiokande experiment places limits on such a neutrino flux \cite{Desai:2004pq}. Neutrinos may be produced either as the direct products of dark matter annihilations or through the decays of bottoms, charms or taus produced in the annihilations.  If the annihilations produce other particles such as muons or light mesons, they will lose most of their energy to interactions in the sun before decaying.  Thus the neutrinos produced have too low an energy to be detected \cite{Hooper:2008cf}.  For the kinetic mixing model of the previous section, the dark matter will annihilate into dark photons; for $\mAp < 2.5 \GeV$ these dark photons cannot decay into charms, taus, or heavier species, and decays to neutrinos are suppressed by the mass of the $Z$, so these constraints can be evaded.  Furthermore, for dark matter masses near the light end of our range, $\lesssim 3$ GeV, these bounds will not constrain the model anyway since the limits cut off rapidly \cite{Fitzpatrick:2010em,Hooper:2008cf}. (Comparison with the limits for elastically scattering dark matter is appropriate as exoDM particles falling into the sun acquire kinetic energies greater than the mass splitting.) In fact, for this light mass range, there may also be a further reduction in the neutrino flux from the sun if the captured dark matter ``evaporates" \cite{Griest:1986yu, Gould:1987ju}.  Thus the light end of the exoDM parameter space may be safe anyway, even if the particle physics model allows annihilations to neutrinos, taus, bottoms, or charms.

\subsection{Annihilation in White Dwarves}

Recent work \cite{McCulloughWD,HooperWD} has brought to light the possibility that dark matter may also be captured inside white dwarves at high rates.  In fact, in certain circumstances, the energy deposited in the star by dark matter annihilations can dominate its luminosity, providing a constraint on the dark matter-nucleon scattering cross section. Because of the large escape velocities of white dwarves ($v_{esc} \sim 10^{-2}$), a keV mass splitting is negligible for a GeV mass DM particle that falls into a white dwarf, so we may consider the scattering as elastic in determining the capture rate.  As emphasized in \cite{HooperWD} with the large dark matter-nucleon scattering cross sections considered in this paper we are in the ``optically thick" regime so the dark matter capture rate does not depend on the size of the cross section (for sufficiently massive white dwarves).  Thus the fact that our cross section is larger than for inelastic dark matter does not affect the limits.  As for inelastic dark matter (iDM), these could rule out exoDM if the density of dark matter in the neighborhoods of certain cool white dwarves was known and found to be large (several orders of magnitude larger than the density in the solar neighborhood).  While these large densities may exist in globular clusters among other places, there are large uncertainties in the actual density of dark matter present.  In particular, it seems impossible to rule out either exoDM or iDM based on current knowledge.  As mentioned in \cite{HooperWD}, one prediction of either of these models is then a lack of cold ($T \leq 4000$ K) white dwarves in environments such as the inner core of the Milky Way or of dwarf spheroidal galaxies.  Conversely, a discovery of such stars would put pressure on exoDM or iDM.

\section{Conclusions}
The nature of dark matter is one of the great cosmological mysteries of our time. The observed complexity in the low energy dynamics of the standard model, with its multitude of cosmologically stable states, makes it plausible that the dark sector could also exhibit similar complexity. Interactions between such a dark sector and the standard model will  generically include exothermic interactions wherein a metastable dark matter particle dumps energy into a nucleus. 

The phenomenology of such interactions in a direct detection experiment is different from that of the conventional elastic scattering between dark matter and a nucleus. The energy deposited in the recoiling nucleus peaks around $\sim \delta \frac{m_\chi}{m_N}$, with a spread around the peak determined by the kinetic energy of the dark matter. The modulation of this kinetic energy spread with the dark matter velocity gives rise to a modulation in the event rate at any given energy, even though the total rate is constant. 

Such a modulation could explain the observations of DAMA while remaining consistent with null observations at a variety of other experiments. Exothermic dark matter preferentially deposits more energy into lighter nuclei than heavier nuclei. This is in stark contrast to endothermic ({\it i.e.} upscattering) dark matter interactions that preferentially scatter off heavier nuclei. The absence of such events in  CRESST-II \cite{CRESSTIICommissioning,CRESSTTalk}, which uses  tungsten (a heavy nucleus), severely constrains these explanations. Other explanations such as form factor dark matter \cite{FormFactor} and momentum dependent dark matter \cite{Chang:2009yt} that also preferentially scatter off of heavy nuclei are also severely constrained by CRESST-II. 

The overall energy scale of nuclear recoils caused by exothermic dark matter is determined by the mass splittings in the dark sector and not by the dark matter kinetic energy. This enables exoDM to fit the DAMA signal with very light WIMPs (2 - 5 keV), while ensuring that the tail of the recoil spectrum remains below threshold in other experiments. This is in contrast to light elastic dark matter  \cite{Fitzpatrick:2010em} explanations of DAMA where the somewhat larger masses needed to accommodate the DAMA signal causes events in low threshold experiments like CDMS Silicon. The absence of such events severely constrains the light elastic dark matter explanations of DAMA.  

Currently, the most stringent constraints on exothermic dark matter arise from the DAMA event rate at $\sim $ keV. Uncertainties in the detector efficiency and energy resolution at these energies makes it difficult to impose stringent limits on this scenario. A better understanding of the response of dark matter detectors at around $\sim$ keV could significantly constrain exothermic dark matter. In particular the installation of new photomultiplier tubes in DAMA may allow an even lower energy threshold \cite{DAMAResults}, leading to additional constraints from the low energy modulated and unmodulated rates. The constraints (or lack thereof) from xenon experiments could be clarified by future measurements of $\Leff$ for liquid xenon, such as those planned by the XENON100 collaboration \cite{Collaboration:2010er}.    

Exothermic dark matter can explain the DAMA modulation using a large cross section without running afoul of other experimental bounds because the nuclear recoils typically occur below the threshold energies of these experiments. Low-threshold experiments and analyses are required to probe this scenario. Studies of low-energy channeling are necessary to determine the true sensitivities of crystal detectors such as CoGeNT to exoDM, and in particular to determine whether the model may simultaneously fit the DAMA and CoGeNT signals. In experiments with heavy nuclei such as XENON10 and CRESST-II, the energy spectrum of exoDM peaks at very low energies, but the event rate is enhanced by the $\sim A^2$ scaling in the cross-section for spin-independent interactions. Because of the very high rate the tail of the spectrum may be visible for sufficiently low thresholds. Experiments with light nuclei such as CDMS Silicon, COUPP and CRESST can be directly sensitive to the peak region for thresholds of $\sim 1 - 2$ keV. Low threshold analyses of these experiments could confirm or rule out this explanation of the DAMA signal.

\section*{Acknowledgments}

We would like to thank Yang Bai, Blas Cabrera,  Savas Dimopoulos, Sergei Dubovsky, David E. Kaplan, Joachim Kopp, Chris McCabe, Matthew McCullough, John March-Russell, Jesse Thaler, Natalia Toro and Jay Wacker for useful discussions. We are particularly grateful to Peter Sorensen for discussions on the cuts employed by XENON10 at low energies. We also thank Jeter Hall for comments of channeling in Ge crystals. S.R. is supported by the DOE Office of Nuclear Physics under grant DE-FG02-94ER40818.  S.R. is also supported by NSF grant PHY-0600465. P.S. is supported by the Stanford Institute for Theoretical Physics and NSF Grant No. 0756174. Fermilab is operated by Fermi Research Alliance, LLC under contract no. DE-AC02-07CH11359 with the United States Department of Energy. 

{\bf Note Added :} While this work was in progress we became aware of interesting work by another
group \cite{Natalia} working on aspects of model building a dark sector at the few GeV scale.  While our results (see Sections \ref{Sec:Constraints} and \ref{Sec:ChanneledIodine}) indicate that the particular parameter space they consider is ruled out by CDMS Silicon and possibly also by XENON10, nevertheless the type of particle physics model they propose could presumably also be used to as a model for the parameter space that we find for exoDM.

\end{document}